# Non-resonant Coherent Amplitude Transfer in Attosecond Four-Wave Mixing Spectroscopy

James D. Gaynor,[1,2] Ashley P. Fidler,[1,2,3] Yuki Kobayashi,[1,4] Yen-Cheng Lin,[1,2] Clare L. Keenan,[5] Daniel M. Neumark,[1,2]* Stephen R. Leone[1,2,6]*

[1]*Department of Chemistry, University of California, Berkeley, CA. 94720, USA.*

[2]*Chemical Sciences Division, Lawrence Berkeley National Laboratory, Berkeley, CA. 94720, USA.*

[3] Currently: *Department of Chemistry, Princeton University, Princeton, NJ. 08544, USA.*

[4] *SLAC National Accelerator Laboratory, 2575 Sandhill Road, Menlo Park, CA, 95024, USA*

[5] Currently: *Department of Chemistry, University of Chicago, Chicago, IL. 60637, USA.*

[6]*Department of Physics, University of California, Berkeley, CA. 94720, USA.*

*\*Authors to whom correspondences may be addressed:* (S.R.L) srl@berkeley.edu; (D.M.N.) dneumark@berkeley.edu

**Abstract**

Attosecond four-wave mixing spectroscopy using an XUV pulse and two noncollinear near-infrared pulses is employed to measure Rydberg wavepacket dynamics resulting from extreme ultraviolet excitation of a 3s electron in atomic argon into a series of autoionizing $3s^{-1}np$ Rydberg states around 29 eV. The emitted signals from individual Rydberg states exhibit oscillatory structure and persist well beyond the expected lifetimes of the emitting Rydberg states. These results reflect substantial contributions of longer-lived Rydberg states to the four wave mixing emission signals of each individually detected state. A wavepacket decomposition analysis reveals that coherent amplitude transfer occurs predominantly from photoexcited $3s^{-1}(n+1)p$ states to the observed $3s^{-1}np$ Rydberg states. The experimental observations are reproduced by time-dependent Schrödinger equation simulations using electronic structure and transition moment calculations. The theory highlights that coherent amplitude transfer is driven non-resonantly to the $3s^{-1}np$ states by the near-infrared light through $3s^{-1}(n+1)s$ and $3s^{-1}(n-1)d$ dark states during the four-wave mixing process.

I. INTRODUCTION

Rydberg wavepackets are nonstationary states composed of a coherently phased superposition of stationary Rydberg eigenstates. These states are typically well characterized in noble gases, lending them well to systematic investigation. Rydberg state lifetimes are expected to scale by $(n^*)^3$, where the effective principal quantum number, $n^* = n - \delta$, accounts for the quantum defect, $\delta$; this follows from established rules for the oscillator strengths of autoionizing states in a Rydberg series [1-3]. Thus, Rydberg series in atomic gases can be used to investigate the evolution of Rydberg wavepackets following the broadband excitation of a coherent superposition of well-defined stationary eigenstates. Recent interest in using Rydberg atoms in quantum simulators by coherently manipulating core electrons of atoms while they exist in excited Rydberg states underscores the importance of precisely measuring coherent interactions among autoionizing Rydberg states [4]. The ability to prepare and manipulate quantum superposition states coherently, affecting their individual components and time-dependent evolution, has long been sought after to develop new quantum technologies [5]. One key aspect of interpreting quantum coherences is precisely understanding how information about the superposition is encoded in the reporting states from which a signal is detected.

Using broadband attosecond light pulses at extreme ultraviolet (XUV) photon energies produced in tabletop optics laboratories, one can generate a coherent electronic wavepacket comprising a superposition of multiple Rydberg states. The evolution of these coherent dynamics may then be followed with attosecond and few-femtosecond (fs) temporal resolution owing to the ultrashort pulse durations that are now routinely available [6]. Attosecond four-wave mixing (FWM) spectroscopy is a powerful new means of measuring ultrafast dynamics in atomic and molecular systems that are excited by such broadband XUV attosecond pulses [7-11]. This FWM technique utilizes a noncollinear beam geometry to generate background free XUV emission signals from phase-matched wave-mixing pathways in a sample. Recent attosecond FWM studies have shown that accurate lifetimes of highly excited individual states are directly obtainable in the time-domain for a few specific cases. For example, the autoionization lifetimes of the $4p^{-1}6d$, $4p^{-1}7d$, and $4p^{-1}8d$ Rydberg states in gaseous Kr were directly characterized by Fidler *et al.* [12] in agreement with frequency-domain literature values. Lin *et al.* [13] measured the few-fs lifetimes of two vibrational levels in gaseous $O_2$ in the $3s\sigma_g$ Rydberg series converging to the $O_2^+$ $c$ state around 21 eV, revealing an interplay between electronic autoionization rate and internuclear distance. In a study of NaCl with $Na^+$ $L_{2,3}$-edge excitation, Gaynor *et al.* [14] identified and characterized the few-fs lifetimes of several atomic-like core-excitons that are highly localized about the $Na^+$ in the ionic solid. In some cases, the accurate time-domain retrieval of dynamics using attosecond FWM spectroscopy can be known, or assumed, to be through resonant transitions between XUV-excited bright states and nearby dark states driven by few-cycle near-infrared (NIR) pulses. However, the FWM process can involve optical transitions between multiple excited states, which may occur resonantly or non-resonantly, and the origin of the FWM signal from each individual state may not arise from the particular observed state. An accurate understanding of time-dependent coherent dynamics in atomic and molecular systems requires careful consideration of each light-matter interaction involved in the FWM experiment [15,16].

Figure 1 illustrates how a broad bandwidth of XUV excited states, such as in a Rydberg wavepacket, is monitored in attosecond FWM spectroscopy. The blue shaded region indicates the coherent excitation of many stationary eigenstates by the broadband XUV light pulse represented by the blue arrow, and the two red arrows reflect consequent time-controlled light-matter interactions between the XUV excited states and nearby dark states that are driven by few-cycle NIR pulses. The two depicted pathways show the possibility for two or more different excited states within the wavepacket to couple directly to the same XUV bright state from which the signal is emitted. As explored in this paper, this mechanism can lead to the amplitudes and timescales from otherwise unexpected states influencing, or even dominating, the dynamics of the FWM emission signals.

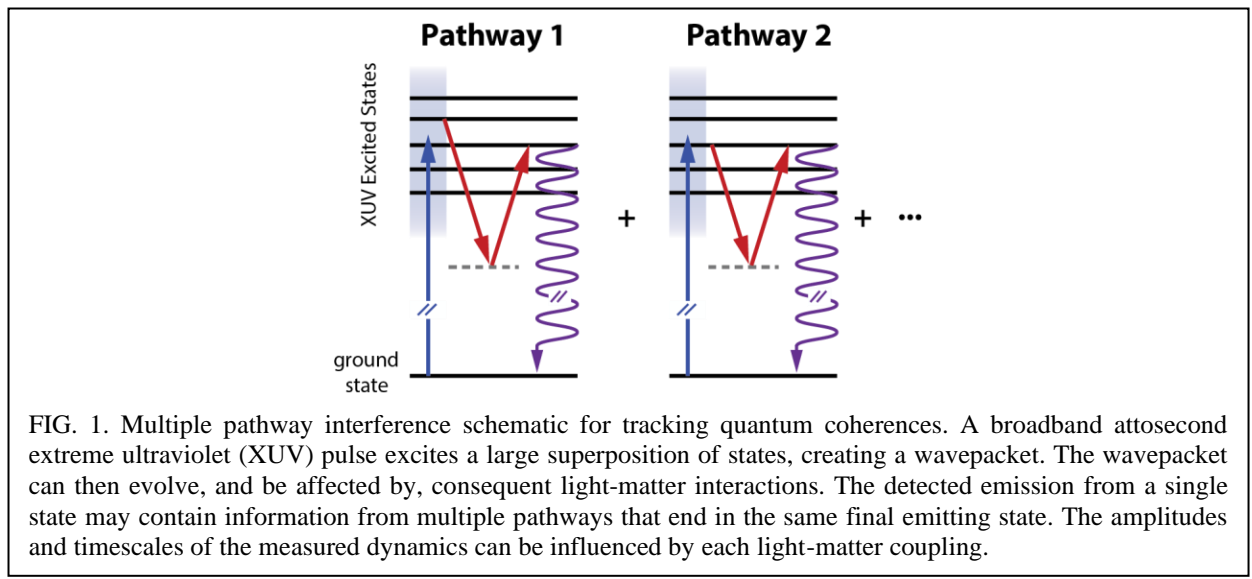

FIG. 1. Multiple pathway interference schematic for tracking quantum coherences. A broadband attosecond extreme ultraviolet (XUV) pulse excites a large superposition of states, creating a wavepacket. The wavepacket can then evolve, and be affected by, consequent light-matter interactions. The detected emission from a single state may contain information from multiple pathways that end in the same final emitting state. The amplitudes and timescales of the measured dynamics can be influenced by each light-matter coupling.

In this work, attosecond FWM spectroscopy is used to study Rydberg wavepacket dynamics arising from single 3s electron excitations in gaseous Ar, accessing the $3s^{-1}3p^6$ $np$ ($^1P$) series of autoionizing Rydberg states that converge to the $3s3p^6$ ($^2S_{1/2}$) level of $Ar^+$ at 29.24 eV [17,18]. A broadband attosecond XUV pulse initiates a coherent superposition of these $3s^{-1}np$ Rydberg states ($n \geq 4$), creating the Rydberg wavepacket. Then, two noncollinear few-cycle NIR pulses probe the evolution of the Rydberg wavepacket. The light-matter interactions complete a phase-matched wave-mixing process, producing an XUV-emission from each of the $3s^{-1}np$ Rydberg states that is spectrally and temporally resolved. Thus, the wavepacket dynamics are projected onto each $3s^{-1}np$ Rydberg state emission. As illustrated in Figure 1, the observed time-dependent FWM emission from each Rydberg state contains information about how the amplitudes of the multiple states in the initial Rydberg wavepacket are transferred through the second and third light-matter interactions driven by the NIR pulses.

A comprehensive analysis of coherent superpositions created by the excitation of an Ar $3s^{-1}np$ Rydberg wavepacket around 29 eV is presented. The experimental measurements are fitted to a wavepacket model to extract the amplitude coefficients of the coherent superposition responsible for the time-dependent FWM emission from individual $3s^{-1}np$ Rydberg eigenstates. Calculations using the time dependent Schrödinger equation support the measurements and wave packet modeling to elucidate the coherent amplitude transfer that occurs during the few-cycle NIR pulse interactions, resulting in the set of coherent oscillations observed in the detected individual state emissions. The results indicate that the coherent decay signals from $3s^{-1}np$ states are dominated by their coupling to longer-lived $3s^{-1}(n+1)p$ states created by the XUV excitation pulse, and that this coupling occurs via non-resonant amplitude transfer through the optically dark $3s^{-1}(n+1)s$ and $3s^{-1}(n-1)d$ states driven by the NIR pulses. Moreover, each emitting $3s^{-1}np$ state displays strong quantum beating, primarily between itself and the $3s^{-1}(n+1)p$ state, with minor components from other states. Overall, the Rydberg series measurements reported here act as a vehicle for discussing the role of non-resonant coherent amplitude transfer in attosecond FWM spectroscopy.

## II. RESULTS AND ANALYSIS

### A. Attosecond Four-Wave Mixing Spectra of Ar Rydberg Wavepackets

The apparatus and technique have been described previously [12,14]. Briefly, the output of a Ti:sapphire laser (Femtopower, 1 kHz repetition rate, 1.7 mJ/pulse, 22 fs, 780 nm) is spectrally broadened in a stretched hollow core fiber compressor (Few-Cycle Inc.) and temporally compressed using a combination of seven chirped mirror pairs (Ultrafast Innovations, PC70), fused silica wedge pairs, and a 2 mm thick ammonium dihydrogen phosphate (ADP) crystal. This yields pulses of sub-6 fs pulse durations with spectra spanning 550-950 nm and ~600 μJ/pulse energies. The beam is then split by a 75:25 (R:T) beam splitter in a Mach-Zender interferometer to separate the NIR driving pulse used to create the XUV pulse by high harmonic generation (HHG) from the pulse used to generate the two noncollinear NIR beams.

Attosecond pulses of XUV radiation in the 25-45 eV range are produced by HHG in Kr (~4 Torr backing pressure). A 150 nm thick Al foil attenuates the co-propagating NIR driving field from the newly generated XUV pulse. The transmitted XUV pulse is refocused by a gold-coated toroidal mirror through an annular mirror into a 1 mm pathlength gas cell for the sample, in which Ar flows out of the laser entrance and exit pinholes at a backing pressure of 14 Torr. The remaining NIR pulse in the other arm of the Mach-Zender interferometer is delayed relative to the XUV beam using a piezoelectric translation stage, and this beam is then further split with a 50:50 beam splitter to create the two NIR pulses for use in the wave-mixing experiment. The relative temporal delay between the two NIR pulses is further controlled using a second piezoelectric translation stage. The two NIR beams are focused and routed through the vacuum chamber to the sample cell to overlap with the XUV pulse spatially and temporally. The two NIR beams are vertically arranged, one above and one below the XUV beam. The interaction of all three beams with the sample produces the phase-matched background free FWM signal.

The NIR pulse envelope is estimated to be 5.6 fs from the rise time of the Ar $3s^{-1}4p$ autoionization signal measured by attosecond transient absorption (ATA). The emitted XUV signals are filtered using another 150 nm Al foil, then spectrally dispersed in the horizontal plane with a flat field grating and recorded using an X-ray CCD camera. In addition to the background free wave-mixing signals, there are ATA signals that co-propagate with the XUV beam. The wave-mixing signals can be isolated from the ATA signal and residual XUV beam, and then further optimized, using a vertically translatable camera mount that allows for the unwanted ATA signals and XUV beam to be translated off of the CCD imaging area.

The transient FWM spectrum of the Ar $3s^{-1}np$ Rydberg wavepacket is shown in Figure 2(a) with temporal lineouts shown as solid lines in Figure 2(b). Time-dependent emissions from the $5p$-$9p$ Rydberg states appear as horizontal features in Fig. 2(a). Signatures from the coherent superposition of Rydberg states are prominent in the FWM signals, manifested as strong intensity oscillations in the emissions from each $3s^{-1}np$ Rydberg state. Clear variations exist in the coherent oscillation periodicity and modulation depth for $3s^{-1}np$ emissions of different principal quantum number. In general, the oscillations have a larger modulation depth and a lower frequency as $n$ increases. The decay times of the $3s^{-1}np$ emissions increase with $n$, in agreement with the well-known trend in autoionization lifetimes. However, the measured decay time of each $3s^{-1}np$ emission is substantially longer than expected from frequency-domain linewidth measurements [17,19,20] and calculations [21] found in the literature for that state. For example, the $3s^{-1}5p$ and $3s^{-1}6p$ state linewidths obtained from synchrotron photoabsorption experiments are 28.2 meV and 12.6 meV, respectively, corresponding to lifetimes of 23.3 fs and 52.2 fs [17,19]. For comparison, single exponential decay functions with the decay rate corresponding to the frequency-domain line widths for the $3s^{-1}5p$ and $3s^{-1}6p$ Rydberg states are shown in Figure 2(b) as dashed blue and dashed red lineouts, respectively. The experimental temporal lineouts of the $3s^{-1}5p$ (blue) and $3s^{-1}6p$ (red) states in Figure 2(b) decay on ~50 fs and ~150 fs timescales, respectively.

The discrepancy between the frequency-domain linewidths and the time-domain FWM decays in Figure 2(b) along with the pronounced quantum beating, conveys that the observed Ar 3s$^{-1}$np Rydberg series FWM emissions exhibit more complex dynamics than a one-to-one mapping of the initial Rydberg wavepacket excited by the attosecond XUV pulse. Previous work [12] has shown this one-to-one mapping can occur when the NIR interactions are resonant with XUV-dark states during V- or Λ-type FWM pathways. To better understand the detected FWM emission from the Ar 3s$^{-1}$np Rydberg wavepacket, the features in Figure 2 were decomposed into the constituent eigenstates in the superposition by determining their relative amplitudes and phase.

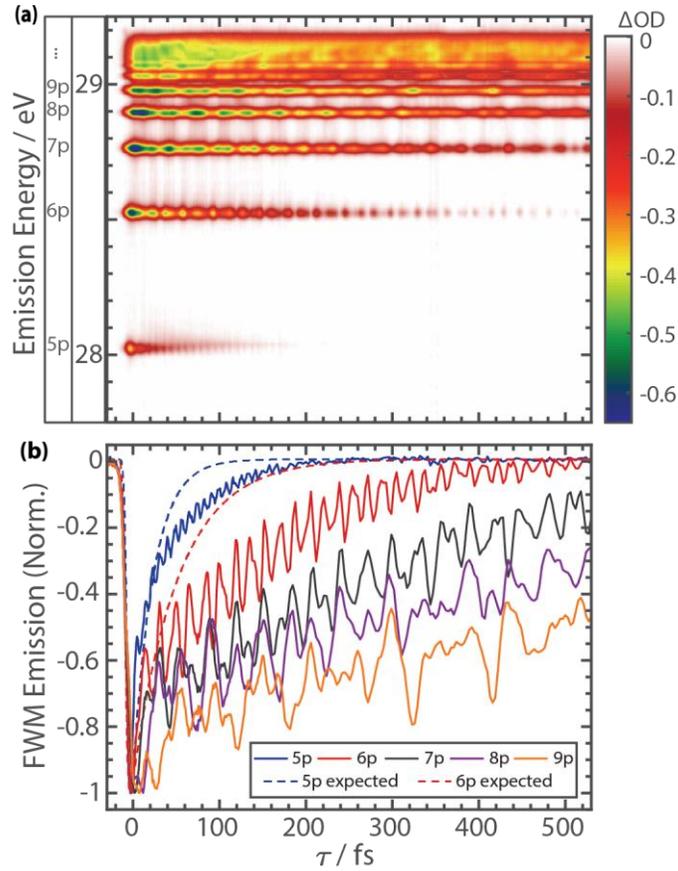

FIG. 2. Strong coherent oscillations in transient four-wave mixing spectra of Ar 3s$^{-1}$np Rydberg series. The transient four-wave mixing spectra of the autoionizing Ar 3s$^{-1}$np Rydberg series (a). Temporal traces taken at the peaks of various Rydberg state emissions to see the temporal oscillations more clearly (b). The solid lineouts in (b) correspond to experimentally measured traces while the dashed blue and dashed red traces are simulated single exponential decay functions with the decay rates specified from frequency-domain linewidth measurements of 23.3 fs for the 5p and 52.2 fs for the 6p, respectively.

Analysis of the 3s$^{-1}$5p emission is presented in Figure 3(a-d) and the 3s$^{-1}$6p emission is shown in Figure 3(e-h). The decay of the $n$th Rydberg eigenstate is represented as $\psi_n(\omega_n, \Gamma_n, t) = -ie^{i\omega_n t + \frac{\Gamma_n}{2}t}$ while $\Psi(t) = \sum_n A_n \times \psi_n(\omega_n, \Gamma_n, t)$ is the coherent superposition of all Rydberg eigenstates. Here, $\omega$ is the Rydberg state energy, $\Gamma$ is the autoionizing linewidth, and $A_n = a_n e^{i\phi_n}$ are the complex superposition coefficients with amplitudes $a_n$ and phase $\phi_n$. The emitted FWM signal reports on the dynamics of the Rydberg wavepacket, and the intensity of the emitted electric field during the FWM process is detected experimentally. The experimental data are fit to $|\Psi(t)|^2$ using this model; to fit the FWM emission traces, the Rydberg state energies were initially set to literature values and allowed to float by ± 10 meV to match more closely the 11 meV experimental resolution of our spectrometer. The fits in Figure 3 were obtained by including the $4 \leq n \leq 10$ members of the $3s^{-1}np$ series. See Supplemental Material for further information on the fitting procedure.

The emitted FWM signal reflects the dynamics of the coherent superposition resulting from the initial XUV-prepared Ar $3s^{-1}np$ Rydberg wavepacket being re-shaped from two NIR light-matter interactions. It is important to note the difference between the initial Rydberg wavepacket prepared by the XUV pulse and the coherent superposition that generates the FWM emission signal. The detected coherence signal results from probing the initial XUV-prepared Rydberg wavepacket with the two NIR light-matter interactions that project part of the initial Rydberg wavepacket onto each final Rydberg state, which then emits a FWM signal.

As shown in Figure 3(a) and 3(e), very good agreement between the model and the experimental data is achieved for the $3s^{-1}5p$ and $3s^{-1}6p$ emissions, giving $R^2$ goodness of fit parameters of 0.99 and 0.96, respectively. The emitted coherence dynamics observed in the $3s^{-1}5p$ and $3s^{-1}6p$ emissions are then elucidated by plotting the $a_n$ coefficients for all eigenstates in Figure 3(b) and 3(f). Interestingly, emission from the $3s^{-1}5p$ state is dominated by the character of the $3s^{-1}6p$ state; similarly, the $3s^{-1}6p$ emission is dominated by the $3s^{-1}7p$ state. This trend also holds for the $3s^{-1}7p$ emission, although the fit is worse for higher $n$ Rydberg states, as neighboring states are energetically closer; see Supplemental Material for 7p fitting.

The FWM intensity oscillations characterizing the coherent superposition are directly analyzed by subtracting the multi-exponential decay that arises from the fits as the diagonal $|A_n \psi_n|^2$ terms. The experimental coherence signatures of the $3s^{-1}5p$ and $3s^{-1}6p$ emissions are isolated in Figures 3(c) and 3(g), along with their Fourier transforms in 3(d) and 3(h), respectively. These signatures reflect the cross terms of the type $(A_n\psi_n)(A_m\psi_m)$ where $n \neq m$. The most intense oscillations of the *np* emission – e.g., peak i at 0.49 eV and peak ii at 0.24 eV in Figures 3(d) and 3(h), respectively – occur at the beat frequency corresponding to the energy difference between the *(n+1)p* and the *np* Rydberg states. These respective energy differences are consistent with the 0.50 eV separation of the $3s^{-1}5p$ and $3s^{-1}6p$ states and with the 0.24 eV separation of the $3s^{-1}6p$ and $3s^{-1}7p$ states as measured in our spectrometer.

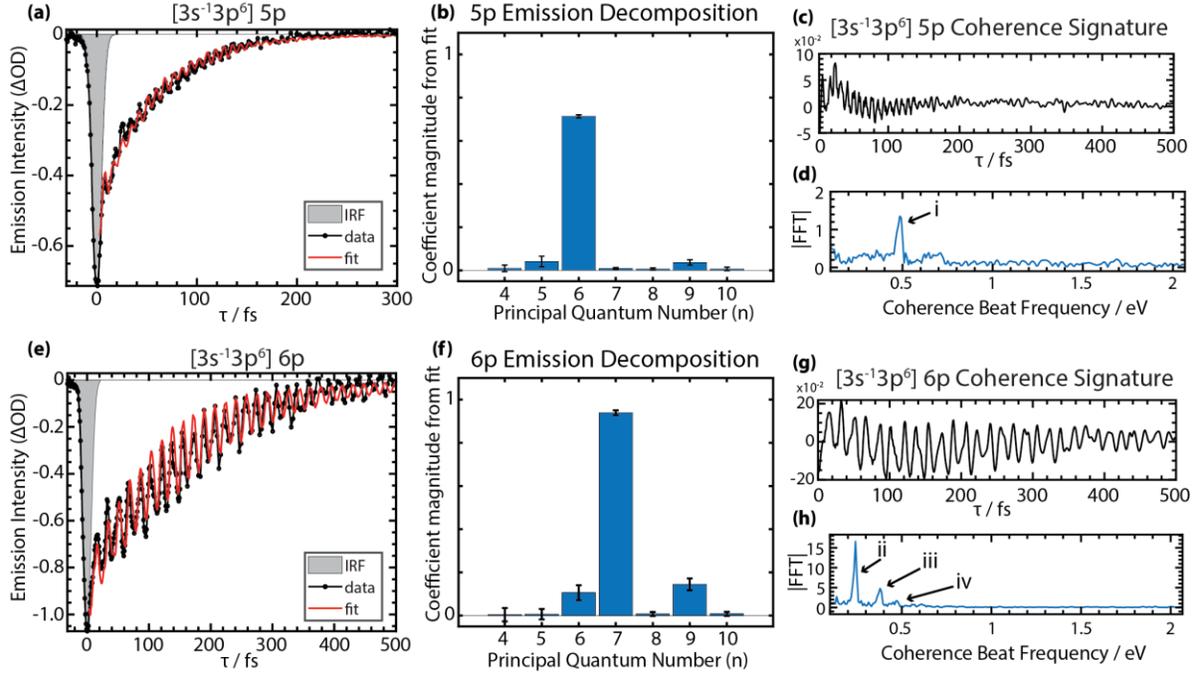

FIG. 3. Extracting the $3s^{-1}np$ state amplitudes composing the Rydberg wavepacket through four-wave mixing emission. The wave-mixing signal of the $3s^{-1}5p$ signal is analyzed in (a-d) and the $3s^{-1}6p$ signal is analyzed in (e-h). The signals are first fit to the wavepacket model described in the text (a,e). The coefficient amplitudes $a_n$ of each component Rydberg state in the wavepacket is plotted as obtained from the fit (b,f) with the error bars representing the 95% confidence interval of the amplitudes. The exponential decay components from the fit are subtracted to isolate the coherent oscillations in the wave-mixing signals (c,g) and then the oscillations are Fourier transformed to identify the primary coherence components in the signals (d,h). Peaks labeled i-iv are discussed further in the text.

It is clear from the analysis in Figure 3 that the NIR wave-mixing pulses drive pathways that transfer amplitude from the *(n+1)p* states of the Rydberg wavepacket to the *np* states that produce FWM emission. There are additional superposition components of smaller magnitude in Figures 3(d) and 3(h). Peak iii appears at 0.38 eV in good agreement with the experimentally measured 0.37 eV energy difference between the $3s^{-1}6p$ and $3s^{-1}8p$ Rydberg states. Although weak, peak iv appears at 0.47 eV and is assigned to interfering contributions between the $3s^{-1}5p$ and $3s^{-1}6p$ coherence and the $3s^{-1}6p$ and $3s^{-1}9p$ coherence, both of which are expected at 0.45 eV. As discussed below in Figure 4, consideration of oscillator strengths between XUV-bright and XUV-dark states that are all far off-resonant with the NIR pulses provide a rationale for why certain states contribute more than others in this specific case.

### B. Calculated Oscillator Strengths: A Quantum Roadmap

In this section, the experimentally observed coherence dynamics are explored theoretically to understand the physical origins of the coherence signatures discussed above with the goal of understanding more clearly how the NIR pulses complete the FWM pathways to produce the emitted signals. To achieve this, the oscillator strengths between the bright and dark states are calculated (Figure 4) and the time-dependent emissions of an Ar $3s^{-1}np$ Rydberg wavepacket are simulated and analyzed (Figure 5). The energies of the Ar ground state, $3s^{-1}np$ ($4 \leq n \leq 9$) states, $3s^{-1}ns$ ($4 \leq n \leq 9$) states, and $3s^{-1}nd$ ($3 \leq n \leq 9$) states are calculated using the Cowan atomic structure code [22]. See Supplemental Material for further details on electronic structure calculations. The oscillator strengths between the XUV-allowed $3s^{-1}np$ states and the XUV-forbidden $3s^{-1}ns$ and $3s^{-1}nd$ states are shown in Figure 4(a) through the relative size of the colored circles. For example, the pink circle in the bottom left reflects the oscillator strength for the transition between the $3s^{-1}5p$ and $3s^{-1}5s$ states and is calculated as 0.494. Similarly, the calculated oscillator strength for the $3s^{-1}5p$ to $3s^{-1}6s$ transition is 0.301. Table I lists the oscillator strengths for the principal transitions shown in Figure 4(a). The oscillator strengths for the other transitions not listed in the table are typically orders of magnitude smaller; see Supplemental Material for complete list of calculated oscillator strengths.

TABLE I. Calculated oscillator strengths for NIR-driven principal transitions (arb. units)

| $n$p | ($n$)s | ($n+1$)s | ($n-2$)d | ($n-1$)d |
|---|---|---|---|---|
| 5p | 0.494 | 0.301 | 0.049 | 0.238 |
| 6p | 0.637 | 0.429 | 0.103 | 0.305 |
| 7p | 0.776 | 0.557 | 0.158 | 0.368 |

Collectively, Figure 4(a) and Table I show that the largest oscillator strengths are for transitions where first the angular orbital momentum quantum number reduces by one (e.g., 7p to 7s in the first column), followed by strong oscillator strengths for transitions where the principal quantum number also changes (e.g, 6p to 7s in the second column). A representative pathway connecting $3s^{-1}7p$ character with the $3s^{-1}6p$ state is pointed out with arrows and labels in Figure 4(a) and with red curved arrows in Figure 4(b). This pathway demonstrates that the $3s^{-1}7p$ excitation can connect to the $3s^{-1}7s$ dark state with the first NIR interaction, followed by the transition to the $3s^{-1}6p$ bright state with the second NIR interaction. In this way, Figure 4(a) serves as a map of the quantum transition pathways that result in the observed coherence signatures and the extended decay times of the Ar $3s^{-1}np$ Rydberg states in Figure 2. Specifically, the results suggest that the $(n+1)s$ and $(n-1)d$ states are the dominant XUV-forbidden states connecting the $np$ and $(n+1)p$ states during the FWM process. When the Rydberg wavepacket dynamics are projected back onto the $3s^{-1}np$ states by the XUV±NIR$_1$±NIR$_2$ pulse sequence, the enhanced $(n+1)p$ character extends the coherent wavepacket dynamics observed in the FWM signal decay time beyond what is expected from the individual $3s^{-1}np$ Rydberg states. These calculations are consistent with the experimental observations shown in Figure 3. Interestingly, as the energy level diagram in Figure 4(b) shows, the observed FWM transition pathways are highly non-resonant for the NIR pulses. The energy separations between identified states involved in the wave-mixing pathways span 0.05 eV-0.35 eV, whereas the NIR pulse spectrum spans 1.23 eV – 2.25 eV.

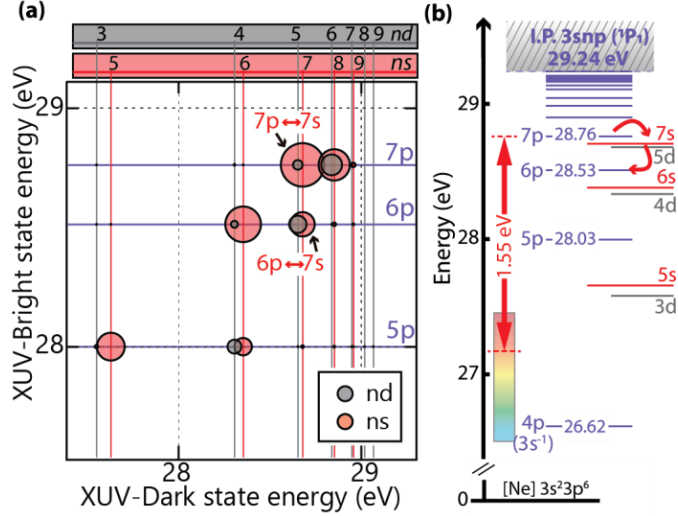

FIG. 4. Calculated oscillator strengths and energy level diagram for XUV-bright and XUV-dark states. The energies of the $3s^{-1}np$, $3s^{-1}ns$, and $3s^{-1}nd$ Rydberg state with respect to the [Ne] $3s^23p^6$ ground state of Ar are shown in (a). The XUV-allowed $3s^{-1}np$ states are plotted on the vertical axis with purple grid lines and the XUV-forbidden $3s^{-1}ns$ and $3s^{-1}nd$ states are plotted on the horizontal axis with pink and grey lines, respectively. The oscillator strengths for the $3s^{-1}np \leftrightarrow 3s^{-1}ns / 3s^{-1}nd$ transitions are represented by the size of the circles; the position of the circle in the two-dimensional plot specifies the transition. An energy level diagram of relevant states in Figure 3 is shown in (b). The center energy and bandwidth of the NIR pulses is shown with respect to the $3s^{-1}7p$ state to highlight the non-resonant condition probed in the experiment (NIR bandwidth is represented by the rainbow-colored box). See text for discussion of the example pathway highlighted in this figure.

To confirm the origin of the time-dependent behavior measured in the FWM experiment, the transient response of the argon atoms is simulated by numerically solving the time-dependent Schrödinger equation (TDSE) where the transition dipole moments between the calculated $3s^{-1}np$, $3s^{-1}ns$, and $3s^{-1}nd$ states are explicitly included. The results are shown in Fig. 5; see Supplementary Material for TDSE simulation details. The TSDE simulation in Figure 5(a) shows prominent coherences and the beat frequencies are identified by Fourier transform analysis in Figure 5(b). The Fourier components from the TDSE simulation highlight the influence of the oscillator strengths calculated in Figure 4(a) through the dynamics. For comparison with theory, the experimental FWM spectrum in Figure 2(a) has also been Fourier transformed over the XUV-NIR time delay ($\tau_1$), which is given in Figure 5(c). We note that the experimental coherence peaks lie on top of strong features at very low frequencies that approach 0 eV due to the underlying exponential decay dynamics. The noise level in Figure 5(c) is greater than the FFT analyses in Figures 3(d) and 3(h) because the underlying multiexponential decay components are not able to be removed globally to enable the enhanced resolution FFT analysis shown in Figure 3.

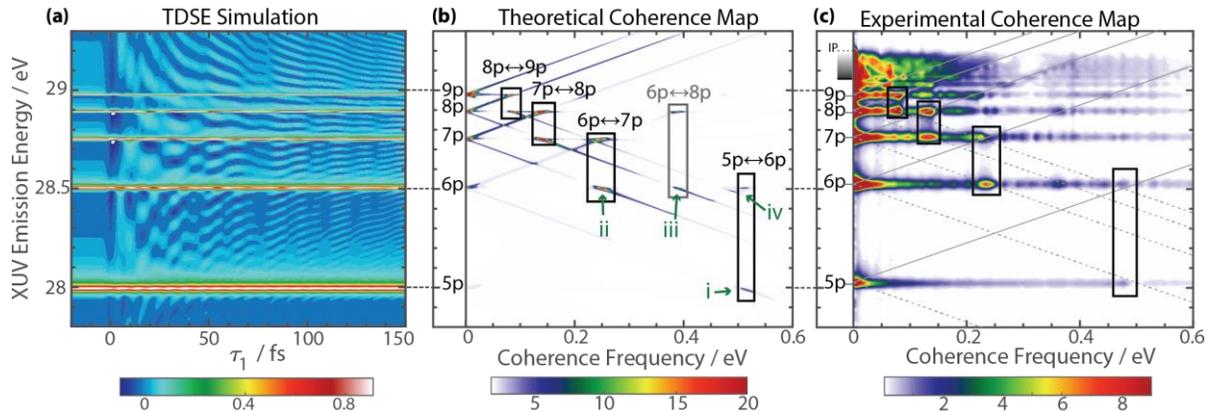

FIG. 5. Theoretical and experimental coherence maps of the Rydberg wavepacket. A simulation of the Ar $3s^{-1}np$ dipole emission by solving the time-dependent Schrödinger equation (a) captures the extended decay times observed in the wave-mixing experiments. The Fourier transform of the simulated signal (b) further matches the coherent oscillatory signatures observed as the Rydberg wavepacket propagates. The experimental wave-mixing spectrum from Figure 2(a) is Fourier transformed for comparison to the result in (b). The solid black boxes in (b) and (c) highlight the retrieved beat frequencies that match the energy differences between the $np$ and $(n+1)p$ Rydberg states, consistent with the wavepacket decomposition trends shown in Figure 3(b) and 3(f). The grey box in (b) shows an additional coherence contribution relevant for the $3s^{-1}6p$ emission, as discussed in the text. Peaks i-iv from the experimental analysis in Figures 3(d) and 3(h) are labeled in green.

The black outlined boxes in Figure 5(b) group the beat frequencies of the coherences involving the $np$ and $(n+1)p$ states. The grey box shows the addition 6p↔8p coherence viewed in the 6p emission. Peaks i-iv, as observed in the experimental analysis of Figures 3(d,h), are labeled in green in the theoretical coherence map of Figure (b), providing theoretical confirmation of the identified coherences and explanation. Peak i is present in the theoretical 5p emission 0.50 eV coherence frequency and 27.99 eV XUV emission energy. Peak ii is observed in the theoretical 6p emission at 0.24 eV coherence frequency and 28.51 eV XUV emission energy. Peak iii appears at 0.38 eV coherence frequency and 28.51 eV. The two expected contributions to peak iv are present in the theoretical 6p emission; the 5p↔6p coherence is observed at 0.50 eV and the 6p↔9p coherence feature appears at 0.46 eV.

III.  DISCUSSION

The coherent nature of the dynamics involving many Rydberg eigenstates is on full display in the presented experiments and accompanying theory, as the oscillatory intensity of the *np* FWM emission extends much longer than expected. For example, the 3s$^{-1}$5p emission has visible oscillations out to ~150-200 fs even though the 3s$^{-1}$6p autoionization lifetime is 52.2 fs. Although higher *n* members of the Rydberg series make up small portions of the amplitude distribution in the fitted FWM emissions, they are still part of the coherent superposition and thus exert influence over its time-dependence. The agreement between the experimental FWM signals and the simulated results shows that the detected coherent wavepacket dynamics are largely driven by the transition dipoles of the NIR coupled states in the excited 3s$^{-1}$*np* Ar system itself, even though there are no XUV-forbidden states (3s$^{-1}$*ns/d*) to complete the wave-mixing pathways lying within the NIR bandwidth. By contrast, previous FWM experiments on the Kr 4s$^2$4p$^5$*nl* autoionizing Rydberg series used NIR-resonant 4p$^{-1}$*np* states to measure 4s$^2$4p$^5$*nl* state lifetimes that matched frequency-domain lifetime predictions [12]. Theoretical investigations of similar off-resonant conditions in attosecond transient absorption spectroscopy also show that the dynamics can be governed largely by transition dipole moments and coherent dephasing rather than by population transfer and relaxation, as occurs when on resonance [23].

The results in this study may be considered through the lens of wavepacket reshaping where coherent amplitude transfer, which occurs through non-resonant NIR light-matter interactions, changes the amplitude coefficients in the superposition of the detected coherence in relation to the initial XUV excited Rydberg wavepacket. These FWM studies of Ar Rydberg wavepackets show that a systematic alteration of autoionization events is possible that is intrinsic to the system itself after a Rydberg wavepacket is launched. As the amplitude distributions show from the wavepacket decomposition in Figure 3, the off-resonant NIR interactions have the effect of redistributing, or transferring, amplitude among the Rydberg eigenstates that compose the coherent superposition – i.e., from the 3s$^{-1}$(*n+1*)*p* states to the 3s$^{-1}$*np* states. Importantly, the nature of the coherent amplitude transfer relies upon the transition dipole moments connecting 3s$^{-1}$*np*, 3s$^{-1}$*ns*, and 3s$^{-1}$*nd* states within the excited state system. It is also significant that the FWM emission decays of the autoionizing *np* states are extended here, rather than shortened, with respect to their literature line-width values. Other time-dependent wavepacket studies have shown detected coherence lifetimes decaying on the timescale of the shorter-lived components in the superposition [24]. Attosecond transient absorption spectroscopy has been used to measure Ar 3s$^{-1}$4p and 3s$^{-1}$5p decay times consistent with frequency domain literature due to NIR coupling with continuum states or, for the 3s$^{-1}$4p state, through resonant dark states accessed by one NIR photon [25]. The non-resonant NIR-coupled pathways in this attosecond FWM experiment show that the transition dipole moments connecting the (*n+1*)*s* and (*n-1*)*d* dark states to the *np* states maintain the coherence in the detected superposition, allowing for the *np* state to be detected through its quantum interference with the *(n+1)p* state long after its autoionization lifetime. The coherent amplitude transfer dynamic reported here sheds new light on important and complex coherent detection dynamics of autoionizing states that are ultimately governed by electron-electron correlations.

There is a longstanding interest in coherently controlling excited state processes with systems ranging in complexity from prototypical atomic Rydberg wavepackets [5,26-30] to mixed electronic, vibrational, and rotational wavepackets in small molecules [24,31,32]. Often, the approach to controlling wavepacket dynamics surrounds the preparation of the coherent superposition using pulse shaping methods with iterative feedback to "customize" the wavepacket composition for particular trajectories. The results in this study do not go as far as coherent control. Frequency-domain pulse shaping methods have been previously used in the attosecond FWM experiment to analyze the NIR-frequency dependence of an attosecond FWM experiment on the Ar $3s^23p^5$ *ns/nd* autoionizing Rydberg series [33]. In this application the few-cycle NIR light-matter interactions were on resonance with optically dark $3s^23p^5$ 4p states, which enabled the isolation of specific wave-mixing pathways from the coherences using the pulse shaper. Other more recent experiments have targeted the $3s^23p^5$ *nf* autoionizing Rydberg series in Ar in which Rydberg wavepacket control is probed through narrower band Raman transitions and using photoelectron-based detection [34]. In contrast, the results presented here provide important insight about the role that non-resonant NIR light-matter interactions can have in coherently transferring amplitude during the attosecond FWM process. More broadly, this work emphasizes the importance of carefully considering the nature of each light-matter interaction and coherence transfer during the probing stages of an initially excited wavepacket.

IV.     CONCLUSION

We have reported attosecond FWM spectra of the Ar $3s^{-1}$ *np* autoionizing Rydberg series around 27-29 eV. This investigation highlights the important role of non-resonant NIR light-matter interactions in probing XUV-excited Rydberg wavepackets. The reported experiments and analysis reveal a mechanism of coherent amplitude transfer arising from the non-resonant NIR light-matter interactions that explain the detected FWM emission coherence decays. Importantly, this work identifies the specific optically dark *(n+1)s* and *(n-1)d* states that are responsible for transferring *(n+1)p* character to the *np* states, from which the FWM emissions are detected. This results in the apparent lifetimes of Rydberg states other than the directly observed individual emissions dominating the FWM coherence decays. The overarching importance of a careful consideration of each light-matter interaction in the attosecond FWM process is emphasized in this work. The insight contributed by this work also has wide ranging implications for current work in quantum information science, where the steps of creation, manipulation, and readout of coherent states are crucial to the performance and robustness of future quantum technologies.

ACKNOWLEDGMENTS

This work was performed by personnel and equipment supported by the Office of Science, Office of Basic Energy Sciences through the Atomic, Molecular and Optical Sciences Program of the Division of Chemical Sciences, Geosciences, and Biosciences of the U.S. Department of Energy at LBNL under Contract No. DE-AC02-05CH11231. J.D.G. is grateful to the Arnold and Mabel Beckman Foundation for support as an Arnold O. Beckman Postdoctoral Fellow. A.P.F. acknowledges support from the National Science Foundation Graduate Research Fellowship Program, and Y.K. and S.R.L. acknowledge NSF grant CHE-1951317 for program support that led to the calculational aspect of the research. Y-C.L. acknowledges financial support from the Taiwan Ministry of Education. Y.K. also acknowledges the support from the Urbanek-Chorodow Fellowship at Stanford University. The authors are grateful to Dr. Kenneth Schafer and Dr. Mette Gaarde for helpful discussions of non-resonant light matter interactions.

J.D.G., A.P.F., Y.-C.L., C.K., D.M.N., and S.R.L. designed the experiments. J.D.G., A.P.F., Y.-C.L., and C.K. performed the experiments. J.D.G. performed the data analysis and modeling. Y.K. provided theoretical calculation and simulations. J.D.G. A.P.F., Y.K., Y.-C.L., C.K., D.M.N., and S.R.L. discussed the results. J.D.G. wrote the paper.